# Comment to: The remote Maxwell Demon as Energy Down-converter

Theophanes E. Raptis[abc]

[a]National Center for Science and Research "Demokritos", Division of Applied Technologies, Computational Applications Group, Athens, Greece.

[b]University of Athens, Department of Chemistry, Laboratory of Physical Chemistry, Athens, Greece

[c]University of Peloponnese, Informatics and Telecommunications Dept., Tripolis, Greece

**Abstract** Taking advantage from a recent analog realization of certain automata via spectral encodings as permutation machines, we reach at an improved version of the previously introduced remotely acting Maxwellian demon in arXiv:1408.3797. A number of alternative encodings for the processing substrate of similar agents is presented together with the new net work gain.

In a recent paper [1], a new mechanism for a Maxwell demon was introduced capable of remotely acting on an external physical system in such a way as to enforce extraction of useful work via a four step thermodynamic cycle. The system to be affected is characterized by a number of units with an asymmetric bistable potential which is then mapped to a set of "binary" symbols denoting lower ("0") or higher ("1") energy states. The demon acts by "copying" a number of these internal states in a memory composed of equal number of subsystems and selectively emits a signal able to cause de-excitation of all subsystems in the higher ("1") states of the external system.

The total cycle is characterized by an amount of waste heat emission $Q_1$ and $Q_4$ for the respective steps while useful work is extracted at the third step as $W$ together with waste heat $Q = W − W_1$ where $W_1$ is the work done by the demon at step one as a measuring signal towards the external system. The efficiency of this scheme has been estimated by the original author as

$$\eta_M = 1 - \frac{Q_1 - Q_4}{W - W_2}$$

We then proceed to show that it may be possible to at least make negligible if not eliminate the waste heat $Q_4$ by an appropriate choice of the internal structure of the proposed demon. Specifically, in [2] we have chosen a particularly simple parallel computing system in terms of the abstract Cellular Automaton (CA) model and showed the existence of alternative representations such that no erasure act takes place thus avoiding as much as possible the production of waste heat at least in the ideal version of such machines. The proposal is in fact much more general than the simple paradigm of a CA and we try to make it as clear as possible with the aid of figure 1. We show there the transcription of the abstract model of a machine with a tape memory as in the case of a universal Turing machine into an equivalent permutation machine. While the original machine "reads" and "writes" symbols from some alphabet, the transcribed version uses cog wheels much like in a "one armed bandit" machine to cause the successive transitions while only a single line of symbols is ever visible through a "slot".

There is then a natural identification of any such cog-wheel either as an oscillator phase or as some other spectral encoding. The method chosen in [2] was based on a particular extension of what is known as a "Manchester code" in the frequency domain so as to guarantee constant spectral density throughout all steps of any computation. Additionally, any such machine would constantly operate with "words" of maximal Shannon entropy $\sim log(b)$ where $b$ the alphabet radix used. We notice in passing that there do exist other possible such representations for abstract machines that are under investigation which may be fitter depending on the task. In particular, taking advantage of the simple inequality $N! > 2^N$ one can also use permutations of a frequency comb to represent for instance, the 256 states of a byte. Given the existence of a lexicographic ordering of irreps for permutations this allows a direct mapping as a Look-Up Table (LUT).

We also notice that the use of a frequency encoding is not necessarily the only option since the constant spectrum can be guaranteed even in the case of a direct wavelength encoding suggesting the case of a cavity with an antenna and a kind of "gateway" buffer for the translation of the received signal into the encoding used. For the particular case of an ideal demon with constant spectral density as a memory we observe that the binary entropy of such a demon remains constant during the process, hence any entropic term $\Delta S$ practically cancels leaving the estimate for the total efficiency as $\eta \leq 1$. The actual significance of the latter observation is that in such a case the constant temperature of the demon is decoupled from that of the machine to be influenced.

In the light of the above observations we believe that there is strong evidence for the necessity to further explore the actual applicability of both the particular application offered by the previous author in [1] and the more general case of spectral machines as generic computing machines.

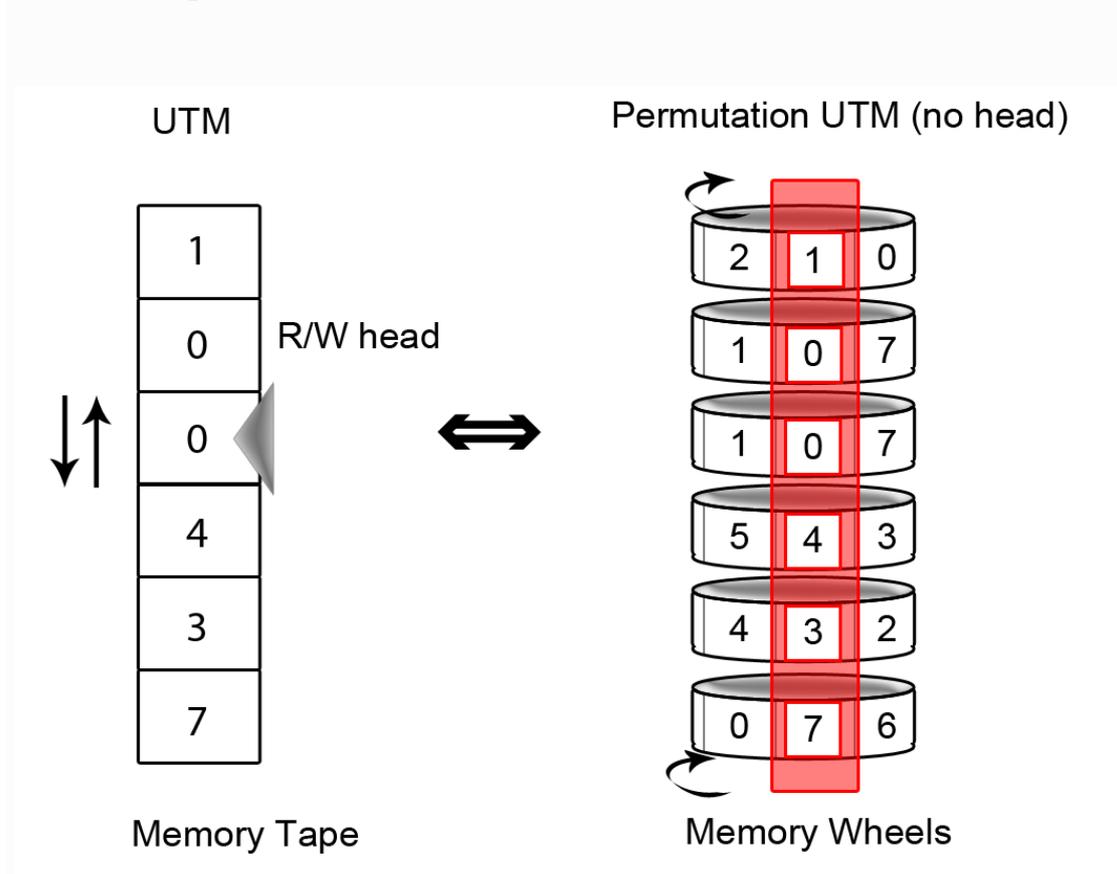

**Fig. 1** Representation of the transcription of an abstract computing machine to its permutation equivalent.